\documentclass[runningheads]{llncs}
\usepackage{graphicx}
\usepackage{xcolor}
\usepackage{hyperref}

\begin{document}
%
\title{Explainability in Practice: Estimating Electrification Rates from Mobile Phone Data in Senegal}

\titlerunning{Explainability in Practice}
%

\author{Laura State\inst{1, 2}\orcidID{0000-0001-8084-5297} \and
Hadrien Salat\inst{3}\orcidID{0000-0003-0958-9715} \and
Stefania Rubrichi\inst{4}\orcidID{0000-0002-5769-8935} \and
Zbigniew Smoreda \inst{4}\orcidID{0000-0002-4047-7597}}
\authorrunning{L. State et al.}
%
\institute{University of Pisa, Pisa, Italy \\
\and
Scuola Normale Superiore, Pisa, Italy
\and
Alan Turing Institue, London, UK
\and
Orange Innovation, Ch\^atillon, France
}
\maketitle              

\begin{abstract}
Explainable artificial intelligence (XAI) provides explanations for not interpretable machine learning (ML) models. While many technical approaches exist, there is a lack of validation of these techniques on real-world datasets.
In this work, we present a use-case of XAI: an ML model which is trained to estimate electrification rates based on mobile phone data in Senegal.
The data originate from the Data for Development challenge by Orange in 2014/15.
We apply two model-agnostic, local explanation techniques and find that while the model can be verified, it is biased with respect to the population density.
We conclude our paper by pointing to the two main challenges we encountered during our work: data processing and model design that might be restricted by currently available XAI methods, and the importance of domain knowledge to interpret explanations.  


\end{abstract}

\section{Introduction}

Explainable AI (XAI) provides techniques to better understand machine learning (ML) models. This is motivated by their lack of transparency, and an increased use of these models in resource allocation problems that critically affect individuals, such as hiring, credit rating, or in public administration.\footnote{\href{https://algorithmwatch.org/en/automating-society-2020/}{https://algorithmwatch.org/en/automating-society-2020/}}
While many XAI methods have been proposed, there is a certain lack of work that uses these methods on real-world data and thus confirms their relevance
\cite{DBLP:journals/corr/abs-1901-04592}. 

In this work, we present a use-case of XAI: 
we train an  ML model to estimate electrification rates in Senegal, and evaluate it using two popular XAI techniques.
The estimation of such socio-economic indicators can support policy planning and is assumed to be a less costly and time-consuming alternative to traditional approaches such as collecting census or survey data. 
Policy planning involves considerable amounts of resources, thus, requires transparency and accountability.
We draw on a dataset of mobile phone data, collected in 2013 and provided during the Data for Development challenge by Orange in 2014/15. We combine it with extracts from the 2013 census in Senegal, and estimate the electrification rate around single cell tower locations.

The contribution of our work is twofold: first, we show how XAI methods can be used to \textit{verify} an ML model, and that our model is biased w.r.t. population densities.
In our case, verifying means showing that the model indeed relies on features that relate to the predicted outcome, as given by domain knowledge.
Thus, we confirm the relevance of XAI techniques.
Second, we point towards two \textit{challenges} of deploying XAI in practice that emerged during this work: pipeline design, and domain knowledge.

The paper is structured as follows: section \ref{sec:background} provides the relevant background, section \ref{sec:data_description} information on the {dataset}, followed by a description of the experiments in section \ref{sec:experiments}. Results are discussed in section \ref{sec:results}, followed by the limitations in section \ref{ref:limitations}. We conclude our paper in section \ref{sec:conclusion}.\footnote{
The code of the project can be found here: \href{https://github.com/lstate/explainability-in-practice.git}{https://github.com/lstate/explainability-in-practice.git}.
}

\section{Background and Related Work}
\label{sec:background}

\subsection{Explainable AI}
\label{sec:xai}

The field of explainable AI can be distinguished along three dimensions: black- vs white-box approaches, local vs global, and model-agnostic vs model-specific~\cite{DBLP:journals/csur/GuidottiMRTGP19}.
The term \textit{white-box} refers to models that are interpretable, or explainable-by-design, while a \textit{black-box} (BB) model is not interpretable, or accessible, e.g., due to intellectual property rights. The majority of used ML models belong to the latter, and it is exactly for those models that we have to design explanation techniques.
These explanations can be distinguished by scale. \textit{Local} techniques explain the prediction for a single data instance (often an individual, in our case a single cell phone tower). Here, prominent approaches are LIME~\cite{DBLP:conf/kdd/Ribeiro0G16} and SHAP~\cite{DBLP:conf/nips/LundbergL17}.
Opposed to this, \textit{global} approaches tackle a full explanation of the system, often by fitting an interpretable surrogate model. One example is the TREPAN algorithm~\cite{DBLP:conf/nips/CravenS95}, building a single decision tree over a neural network.
Last, we differentiate between approaches that work on any (\textit{model-agnostic}) or only one (\textit{model-specific}) model. LIME~\cite{DBLP:conf/kdd/Ribeiro0G16} and SHAP~\cite{DBLP:conf/nips/LundbergL17} are considered model-agnostic, TREPAN~\cite{DBLP:conf/nips/CravenS95} model-specific.
A full survey of approaches can be found elsewhere, e.g.,~\cite{DBLP:journals/csur/GuidottiMRTGP19,DBLP:journals/access/AdadiB18,molnar2019}.
The latter includes a full chapter on interpretable (white-box) models.

\paragraph{LIME (Local Interpretable Model-agnostic Explanations)} ~\cite{DBLP:conf/kdd/Ribeiro0G16}
This method uses a randomly generated neighborhood, weighted according to a distance and a kernel function, to fit a linear regression around the data instance in focus. This regression approximates the decision boundary, and its weights are interpreted as the \textit{importance} of the features. A positive importance pushes the (here) classifier towards the predicted class, while a negative pulls it away and towards one of the other classes. Therefore, both the sign and magnitude of the importance matter.

\paragraph{SHAP (SHapley Additive exPlanations)}~\cite{DBLP:conf/nips/LundbergL17}
Explanations are based on a game-theoretic approach, and provide for each feature its \textit{contribution} towards the predicted outcome. Contributions are not only specific to an instance but also to the (here) class. If we sum over all contributions w.r.t. to a class $C$, and add this value on top of the expected value $E_C(f(x))$, we reach the predicted value of our model $f(x)$, which in turn determines the class (the highest value wins). 
Therefore, the sign and magnitude of a feature contribution matter as well.

\paragraph{Shortcomings LIME and SHAP}
Both methods have some well-known shortcomings. We refer here to a few:
regarding LIME, an open issue is the (mathematical) definition of a neighborhood 
as well as its robustness~\cite{molnar2019,DBLP:journals/corr/abs-1806-08049}.
SHAP, on the other hand, has a well-defined mathematical background. Issues are its computational complexity and (for KernelSHAP) feature dependencies that are not considered.

\subsection{Mobile Phone Data and Electrification}
\label{sec:electrification}

Mobile phones are a rich source of information, providing details about time, length and location of calls and other data.
Combined with the fact that mobile phone penetration is generally high,\footnote{\href{https://ourworldindata.org/grapher/mobile-cellular-subscriptions-per-100-people}{https://ourworldindata.org/grapher/mobile-cellular-subscriptions-per-100-people}}
this opens possibilities for research, public policy, infrastructure planning, etc.
Predicting socio-economic indicators from remotely accessible data is popular, and we observe an interest in using such data in countries of the Global South, where it might be a less costly and time-consuming alternative to traditional approaches such as collecting census or survey data.
Different indicators can be predicted, and they vary based on available data and methods.
Based on mobile phone data, examples are the estimation of socio-economic status and welfare indicators, literacy rate, population densities, and electric consumption~\cite{Blumenstock2015a,Schmid2016,salat_p1}. 
For approaches using additional data sources, an example is the estimation of poverty measures ~\cite{Pokhriyal2017,Steele2017}.
In many cases, these studies are framed within the 17 SDGs.\footnote{\href{https://sdgs.un.org/goals}{https://sdgs.un.org/goals}}
\vspace{0.25cm} \\
In this work, we focus on the estimation of electrification rates using mobile phone data. 
Relevant studies that investigate the relation of electricity and other indicators such as mobile connectivity or volume of visitors in Senegal are~\cite{Houngbonon2021,salat2020impact}.
What is novel to our work is that we do not only estimate electrification rates, but we use XAI methods to understand these estimations and evaluate them.
As such, we present a use-case of XAI, with a focus on the Global South.
A recent survey on XAI projects centered around the Global South~\cite{DBLP:conf/dev/OkoloDV22} showed that while the body of work in the field of XAI is growing rapidly, only 16 of the surveyed papers relate to the Global South (approx. 18,000 papers on the topic of XAI, in the same time span), with only one technical study that has similar focus and design as ours (focus on interpretable poverty mapping)~\cite{DBLP:journals/corr/abs-2011-13563}.

\section{Dataset}
\label{sec:data_description}

\subsection{Mobile Phone Data}
\label{sec:mobile_phone_data_}

We use mobile phone data provided in the form of pre-aggregated call detail records (CDR), which were made available during the second Data for Development challenge launched by Orange in 2014/15.
The original data were collected by Sonatel ({\it Société Nationale des Télécommunications du S\'en\'egal}), which is the leading telecommunication company in Senegal (market share of $65 \%$ in 2013).
CDRs are generated for billing purposes and are proprietary. 

Senegal is a sub-Saharan country, located in the Northern Hemisphere near the Equator, on the West Coast of Africa. It covers 196,712 km$^2$. 
In 2013, the year when the data were collected,
the population count approached 14M.\footnote{\href{https://data.worldbank.org/country/senegal}{https://data.worldbank.org/country/senegal}}
The original dataset contains more than 9 million individual mobile phone numbers, with an hourly resolution. Sonatel anonymized the data, Orange pre-aggregated and processed it further.
The resulting dataset holds (cell) tower-to-tower activity, for calls (including call length) and text messages separately. 
Spatial coverage between call and text message data is different, text message data less available in the Eastern part of Senegal (see {also} appendix, section \ref{sec:data_distribution}).
These areas fall together with areas that are less electrified 
which might suggest a connection between access to electricity, poverty rate, literacy, and text message activity (see~\cite{Houngbonon2021} for a study on the impact of access to electricity).
All details on the dataset can be found here~\cite{DBLP:journals/corr/MontjoyeSTZB14}.

\begin{table}[]
    \centering
    \caption{Features and abbreviations. {The type of an event} (number of calls, call length or number of text messages) will be indicated by subscripts (CN, CL, SN, respectively).}
    \begin{tabular}{ccl}
    te  && number of events \textit{within} Voronoi cell  \\
    && (synonymously used: total events) \\
    out && number of outgoing events\\
    in && number of incoming events \\
    dc && degree centrality \\
    cc && closeness centrality \\
    out/in && ratio of outgoing over incoming events \\
    \end{tabular}
    \label{tab:nw_features}
\end{table}

We process this base data according to the following steps: 
1) building one network per data type, amounting to three networks in total (number of calls, call length and number of text messages). Cell towers form the network nodes, and are labeled according to the electrification rate;
2) extracting each six features per network, amounting to a tabular dataset of a $1587 \times 18$;
3) binning the resulting dataset to create an ordered classification problem of 10 classes;
4) sub-sampling due to an imbalance of the dataset.

Regarding the network construction (step 1), we proceed as follows: we build a directed, weighted communication network from the data, based on the overall activity of 2013, i.e. with no resolution over time. While cell towers form the nodes, edges are created based on the activity in the network. 
Per cell tower (outgoing site in the original dataset) we aggregate over the receiving tower (incoming site in the original dataset) and sum over respective events, for example the number of calls between those two towers, and ignore thereby the time stamp. 
The total number of events per connection, and in the full year 2013, determines the weight of the edge.
This will lead to a full matrix. It has elements on its main diagonal, as calls/messages appear also within a Voronoi cell. Also, it is not symmetric as the number of outgoing and incoming calls/messages between two cell towers is generally not the same.
We repeat this network construction for all three types of data (call numbers, call length and message data). 
Cell tower locations are used to map the activity spatially, we rely on a Voronoi cell tessellation. The electrification rate of each of the Voronoi cells is assigned based on the 2013 census in Senegal.

Regarding the second step, the extracted features are listed in table \ref{tab:nw_features}, they form the final tabular dataset.
The number of events within a Voronoi cell, as well as outgoing and incoming events, are helpful for a general understanding of the activity in the network. 
Centrality measures are added as they are basic measures of communication networks.
Further, centrality measures and the ratio between outgoing and incoming events are inspired by ~\cite{salat_p1}.

Steps 3 and 4 refer already to data preparation and are therefore discussed in section \ref{sec:data_preparation}.

\subsection{Electrification Data}
\label{sec:socio_economic_data}

Electrification rate, population count, and population density originate from the 2013 census in Senegal~\cite{ANSD}. Pre-processing is identical to Salat et al.~\cite{salat2020impact}. The census contains questions about the source of lighting for each household and therefore informs us of stable access to electricity. In the context of this study, stable access means access either to the main power grid or to photo-voltaic systems that benefit from the year-round high solar irradiation in the country. The electrification rate of each census unit ({\it commune}) is given by the ratio between the total number of households with stable access and the total household count. We assume a homogeneous distribution of the population inside each commune, therefore the electrification rate of a Voronoi cell is given by the average electrification rate of all intersected communes weighted by the area of intersection. Salat et al.~\cite{salat2020impact} report that the resulting electrification rates were in good agreement with the fine-grained nighttime lights intensity provided by NOAA~\cite{NOAA}. The supporting shapefile containing the geographic boundaries and population counts at commune level was provided alongside another previous study\cite{salat_p1}.

\begin{figure}[ht]
    \centering
    \includegraphics[width=0.6\linewidth]{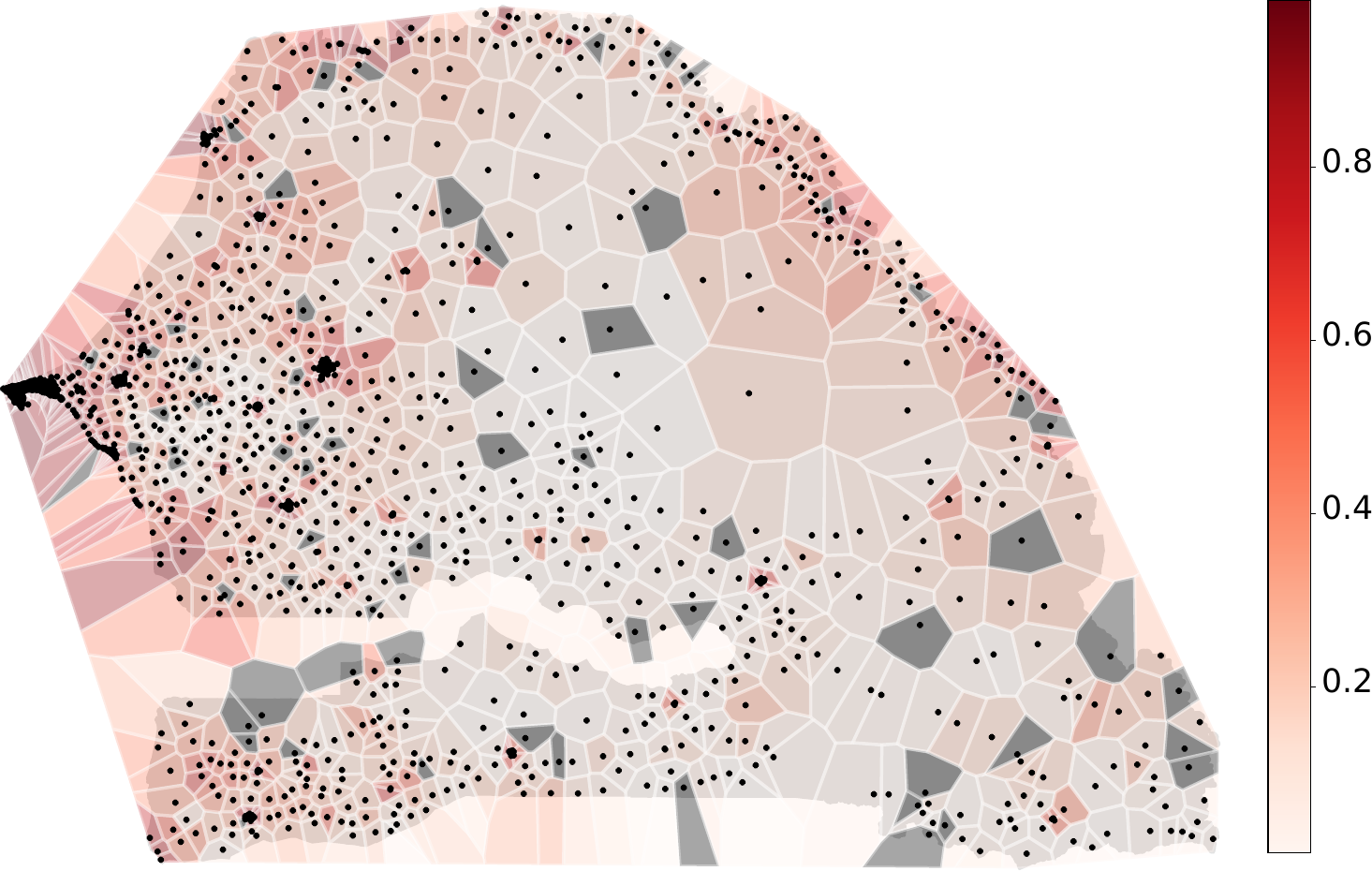}
    \caption{Electrification rate as computed from census data, plotted over the Voronoi cell tessellation. 
    }
    \label{fig:senegal_elec}
\end{figure}

Figure \ref{fig:senegal_elec} shows electrification rates in Senegal. Higher rates are observed around major cities: Dakar, the Capital, in the Western most part of the country, Saint-Louis in the North West, and the culturally significant city of Touba in the center. The electrification rates are also high along the more densely populated Northern/Eastern borders with Mauritania. 
This follows closely the electric grid of Senegal, discussed in~\cite{DBLP:journals/corr/Martinez-Cesena15}.
{We, therefore, observe a correlation between areas with a high electrification rate and a high density of cell towers.}

\section{Experiments}
\label{sec:experiments}

\subsection{Preparation}
\label{sec:data_preparation}

Each row in the pre-processed tabular dataset (corresponding to a Voronoi cell/cell tower) is labeled by its electrification rate (for the pre-processing, see section \ref{sec:mobile_phone_data_}).
We bin the data such that it holds 10 classes, ordered by electrification rate, and with a bin size of $0.1$ ({$[0, 0.1)$ electrification rate for class 0, $[0.1, 0.2)$ electrification rate for class 1, etc.).}
The resulting distribution is skewed towards higher values.\footnote{The ratio between instances in class 9 and the full dataset before subsampling is $imb = 33 \% $.}
We subsample elements of class 9 (electrification rate $[0.9, 1]$), for the training sets only. We sample such that numbers in class 9 are reduced to the average count over all classes. 
Thus, we predict the electrification rate (as class), based on 18 features.
We randomly split the dataset to create a training and a test set for the ML classifiers ({ratio} $7:3$), and split the test set to create a training and a test set for the explanations ({ratio} $7:3$).

\subsection{Models}

\paragraph{Classification} 
We train a set of different ML standard models to estimate electrification rates, the best-performing model will be the basis for all future experiments.
This is based on the assumption that only this model would be used in deployment. However, investigating the explanations for the remaining ML models could also provide valuable insights, and remains for future work.

First, we train a decision tree (DT), a random forest model (RF) containing $100$ decision trees, and an extreme gradient boosting model (XGB) with default parameters.
We use default parameters to ensure better reproducibility.
Additionally, we run a simple test on the following ML models: a logistic regression (LOG), an AdaBoost classifier (ADA), a support vector machine (SVC), a multi-layer perceptron (MLP) and a Gaussian Naive Bayes model (BAY), all using default parameters.
We compute the accuracy, the mean absolute error (MAE), and the ratio between the MAE and its maximum ($MAE_{max} = 9$).

\paragraph{Explanations} We compute local, model-agnostic explanations based on LIME and SHAP. 
A basic assumption of local XAI methods is that it is generally easier to approximate the decision boundary of a complex model locally and thus provide more faithful -- i.e. better -- explanations, than at the global level
\cite{DBLP:conf/kdd/Ribeiro0G16}.
We particularly choose LIME and SHAP for the following reasons: both methods, including their shortcomings (see section \ref{sec:xai}), are well known in the community.\footnote{$12.7M$ downloads of LIME python package, $63M$ downloads of SHAP python package, retrieved on 6th of April 2022 \href{https://pepy.tech}{https://pepy.tech}}
Methods are also easily accessible and rely on interpretable features~\cite{molnar2019}.

As LIME needs to compute some basic data features to generate explanations, it has to be initialized on the explanation training set. 
We use LIME tabular, retrieve the five most important features ($d = 5$), and use default parameters. Explanations are computed once over the given sets.
Evaluations of explanations are computed over the explanation test set.
For SHAP, we use the tree explainer, as we focus on the random forest model, and also default parameters.

\subsection{Urban and Rural Areas} 
We hypothesize that the ML model predicts electrification rates with different accuracy for rural and urban areas, i.e. that the model is \textit{biased} w.r.t to the population density. Therefore, we identify these regions in the test datasets and calculate the disaggregated accuracy values. Further, we compute the disaggregated explanations.
We identify urban regions based on a population density $p > 1000 / km ^ 2$, as previously done in~\cite{salat2020impact}, otherwise as rural. 

\section{Results}
\label{sec:results}

\subsection{Classification}

Best performance is achieved by the random forest model ($acc = 0.516, MAE = 0.972$, see table \ref{tab:results}). 
However, excluding ADA and BAY, all models perform similarly well (extended results can be found in the appendix, section \ref{sec:additional_results_accuracies}).
As the classification is ordered, an MAE around one means that on average, the electrification rate is wrongly estimated by $0.1$, which is acceptable.
Note that while we need a sufficiently high performance of our models to proceed, the focus of this paper is not on model performance.

\bgroup
\begin{table}
    \centering
    \caption{Classification results. The higher the accuracy ($acc$) the better, the lower the $MAE$ the better. Best values are underlined.}
    \begin{tabular}{|c|c|c|c|c|c|}
    \hline
    model & $acc$ & $MAE$ & $MAE$ / $MAE_{max}$ & $acc_{urban}$ & $acc_{rural}$ \\
    \hline
    \hline
         DT &  0.428 & 1.258 & 0.140 & 0.714 & 0.247 \\ 
         RF &  \underline{0.516} & \underline{0.972} & \underline{0.108} & 0.854 & 0.301 \\ 
         XGB &  0.491 & 1.027 & 0.114 & 0.789 & 0.301 \\
    \hline
    \end{tabular}
    \label{tab:results}
\end{table}
\egroup

\subsection{Explanations}

\begin{figure}
    \centering
    \includegraphics[width=0.45\linewidth]{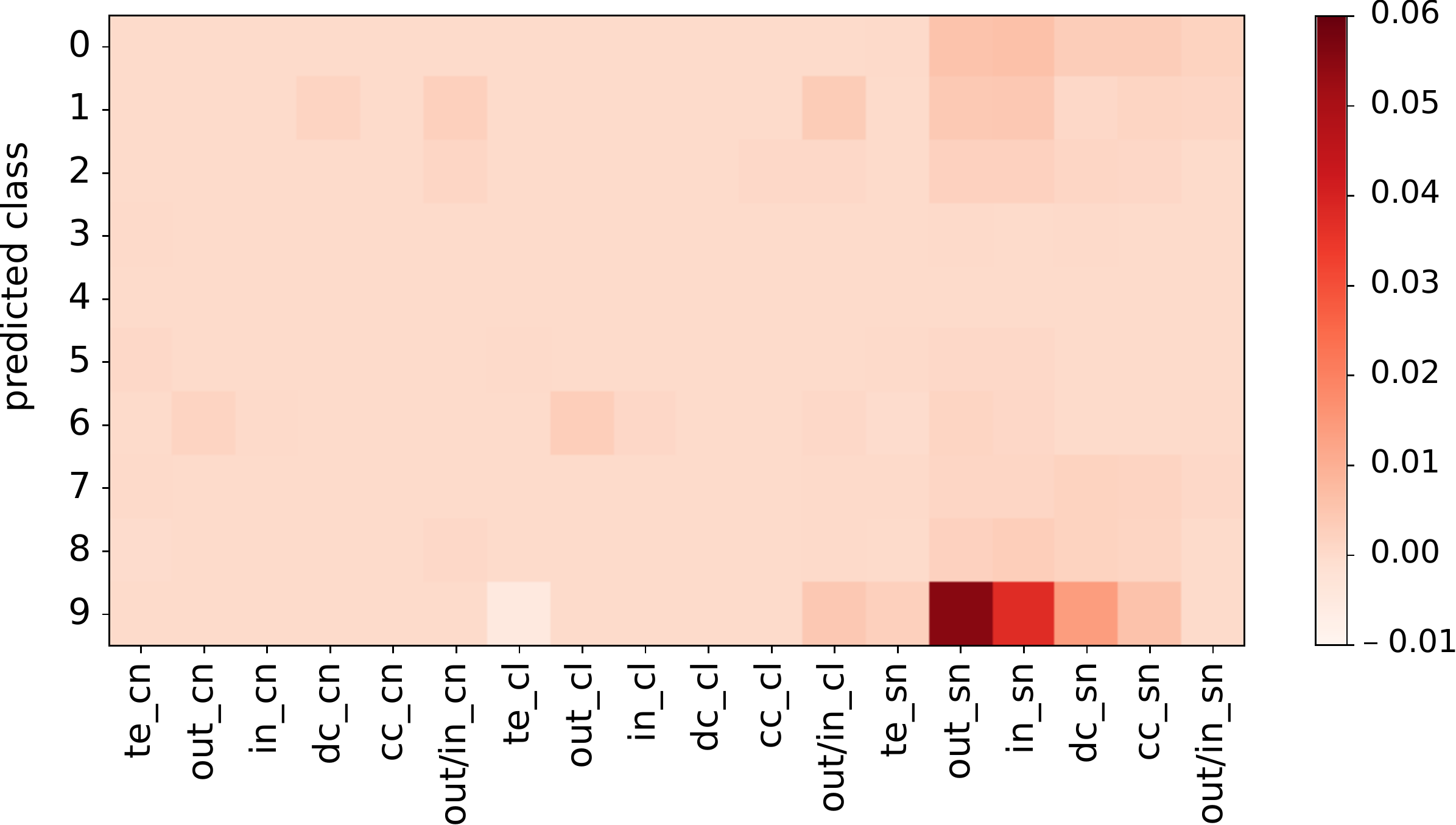}
    \includegraphics[width=0.45\linewidth]{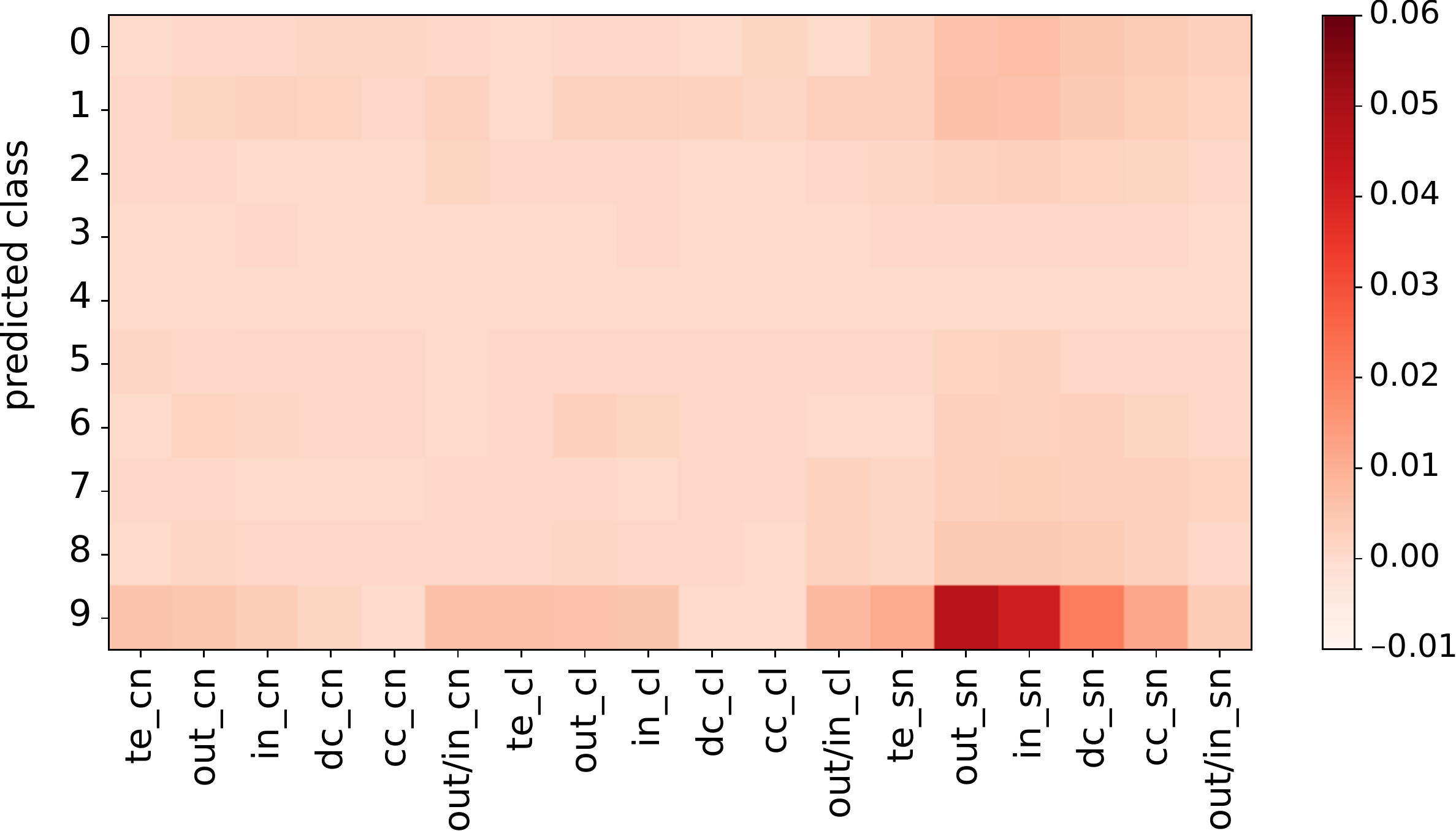}
    \caption{Average importance (left, LIME) or contribution (right, SHAP) of a feature w.r.t. the predicted class. 
    Both plots computed for RF classifier.
    Reminder of notation: table~\ref{tab:nw_features}.
    }
    \label{fig:exp_evaluation}
\end{figure}

We display results in figure \ref{fig:exp_evaluation}, where we plotted the average \textit{importance} (left, LIME) and the average \textit{contribution} (right, SHAP) w.r.t. the \textit{predicted} class.
Both for LIME and SHAP, features based on text message data are highly relevant. This is especially prominent for outgoing and incoming events and the degree centrality and for class 9.
Only in the case of LIME, we observe small negative values (e.g., total events based on call length, for class 9).
The importance, \textit{and} high contribution of text message data for the prediction is in line with the observation that text message activity could be correlated with the electrification rate (see also section \ref{sec:mobile_phone_data_}).
As such, the information provided by this type of data is highly relevant for the model, providing cues to better discriminate. 

\begin{figure}[h]
    \centering
    \includegraphics[width=0.475\linewidth]{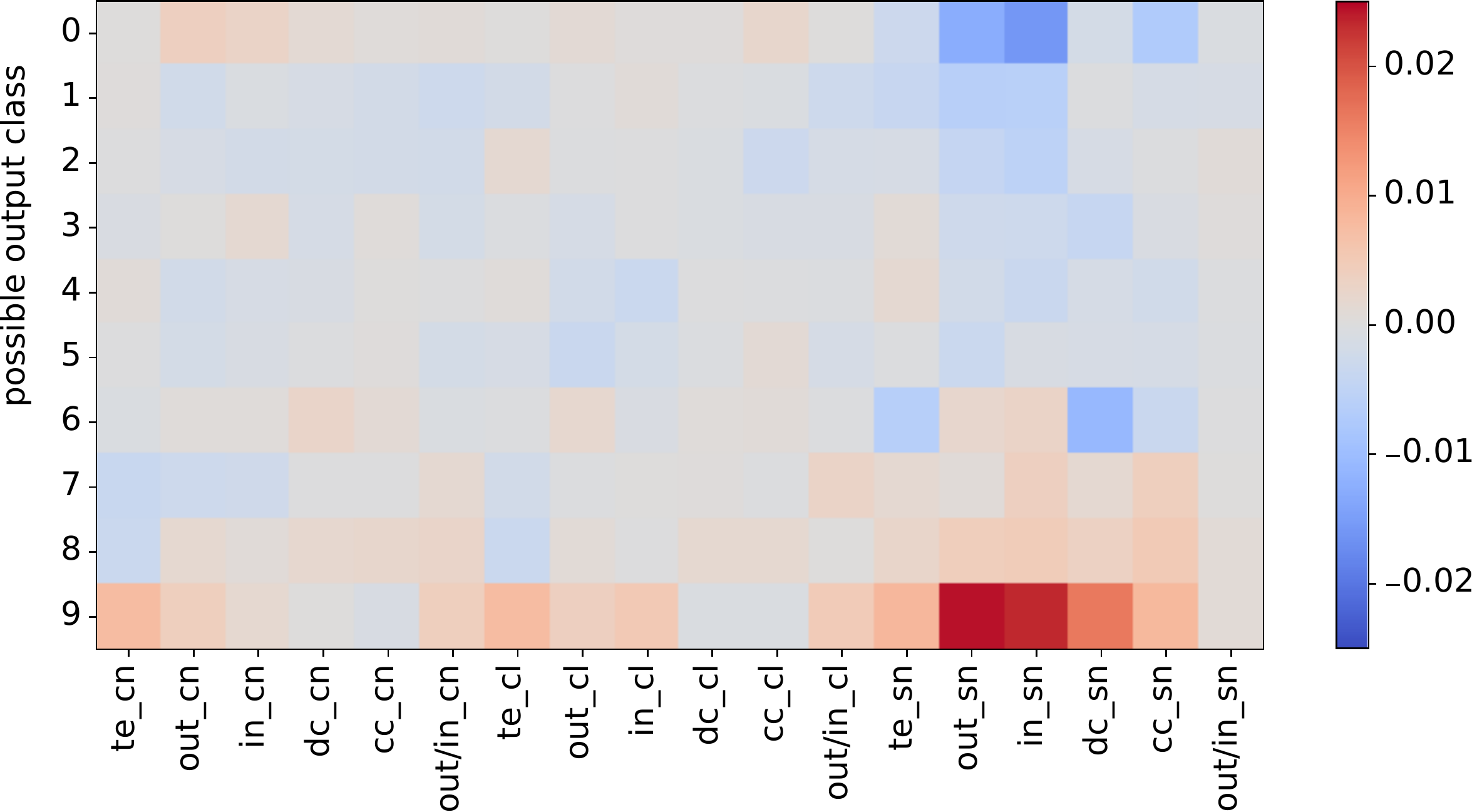}
    \caption{Average contribution of a feature w.r.t. the possible output class.}
    \label{fig:shap_full}
\end{figure}

For SHAP, we also computed the average over feature contributions w.r.t. \textit{all possible} output classes (see figure \ref{fig:shap_full}). 
Why do we observe negative contributions for low classes? 
For a single data instance, features based on text message data generally push the model towards the predicted class (high positive contribution) but at the same time away from the other classes (high negative contribution), together they form the set of possible output classes.
This effect is more prominent for data instances from higher classes. Positive contributions are particularly high for high classes, while negative contributions are high for low classes.
Opposed, in figure \ref{fig:exp_evaluation}, right, for better comparability with LIME, we only plotted contribution w.r.t. to the \textit{predicted} class (and not all possible output classes), thus only positive contributions appear.

Both methods confirm that the model relies on features that indeed relate to the predicted outcome.
While a more in-depth analysis of the model and explanations is needed, this is a first good result.

We note that displayed values are only averages, thus positive and negative relevance values w.r.t. the same feature and class can cancel each other out.
Nevertheless, they provide an informative summary of the model behavior.

\paragraph{Comparing LIME and SHAP}
While both approaches produce some measure of \textit{feature relevance}, the methods are different, thus
the meaning of a feature relevance is not the same (see also section \ref{sec:xai}). 
On one hand, if we study figure \ref{fig:exp_evaluation}, we find only small differences.
SHAP, on the other hand, can give us some additional information, if we consider the full output (see figure \ref{fig:shap_full}).
However, in our particular case, it does entail no surprising insights.
Therefore, while we acknowledge the mathematical basis of SHAP as an advantage, we cannot make a strong argument in favor of it, and against LIME.

\subsection{Urban and Rural Areas}
Disaggregated accuracy values are displayed in table \ref{tab:results}, and in the appendix in section \ref{sec:additional_results_accuracies}.
Excluding the AdaBoost model, the accuracy for rural areas is always lower than the accuracy as computed for the full test set, and the opposite is the case for urban areas.
This difference in accuracy means that the model is \textit{biased} w.r.t. to the population density.

\begin{figure}[ht]
    \centering
    \includegraphics[width=0.425\linewidth]{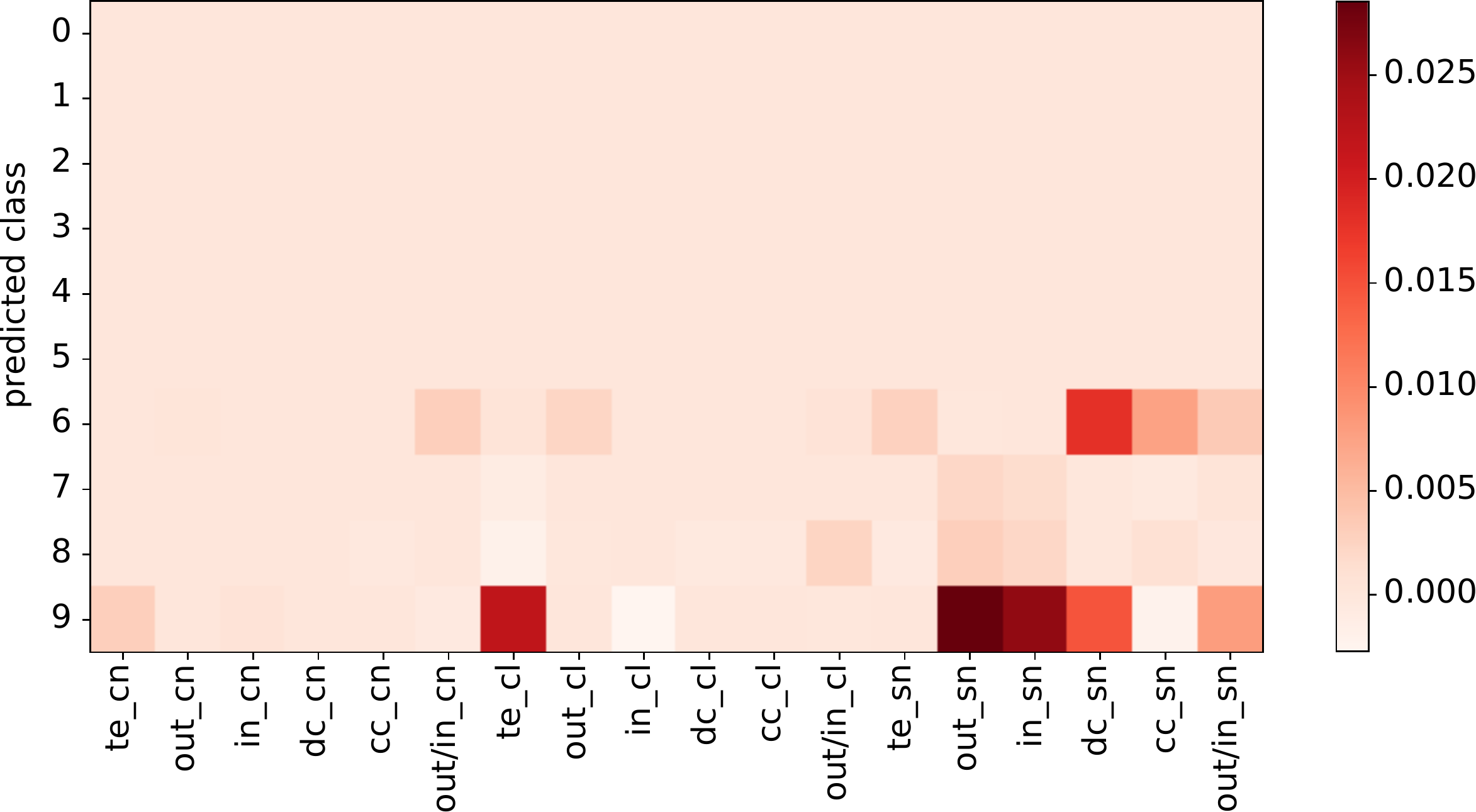}
    \includegraphics[width=0.42\linewidth]{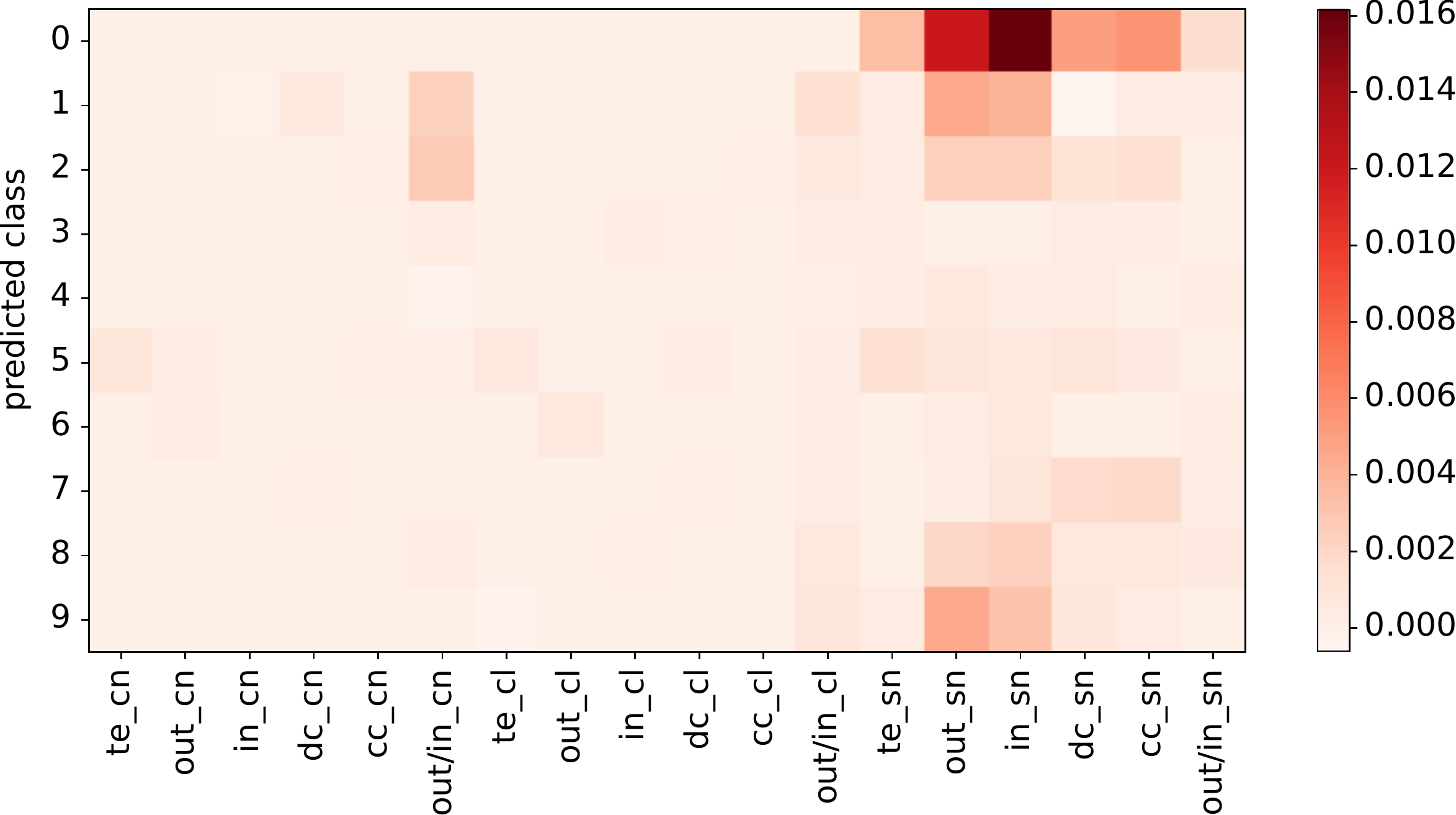}
    \includegraphics[width=0.42\linewidth]{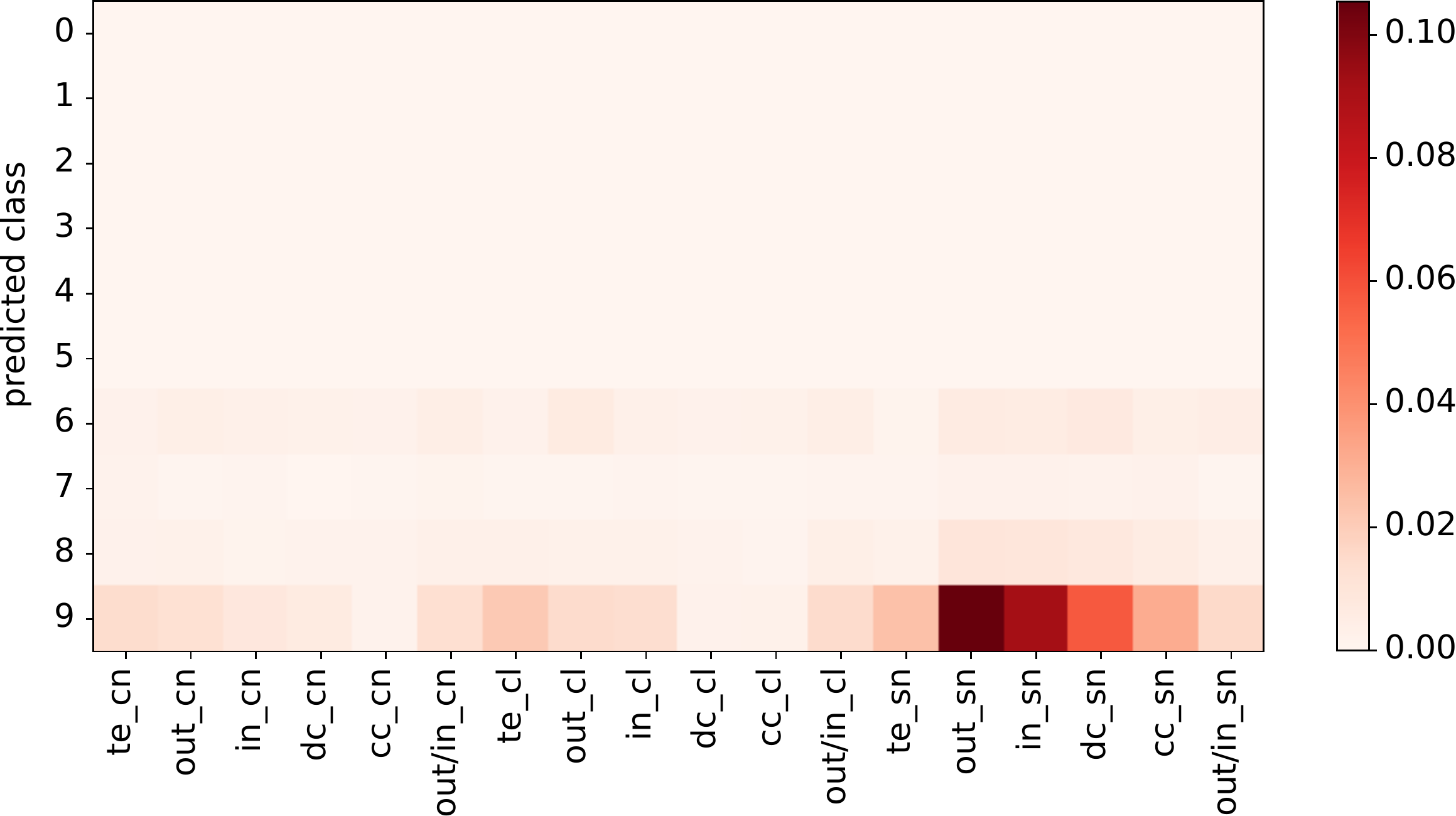}
    \includegraphics[width=0.43\linewidth]{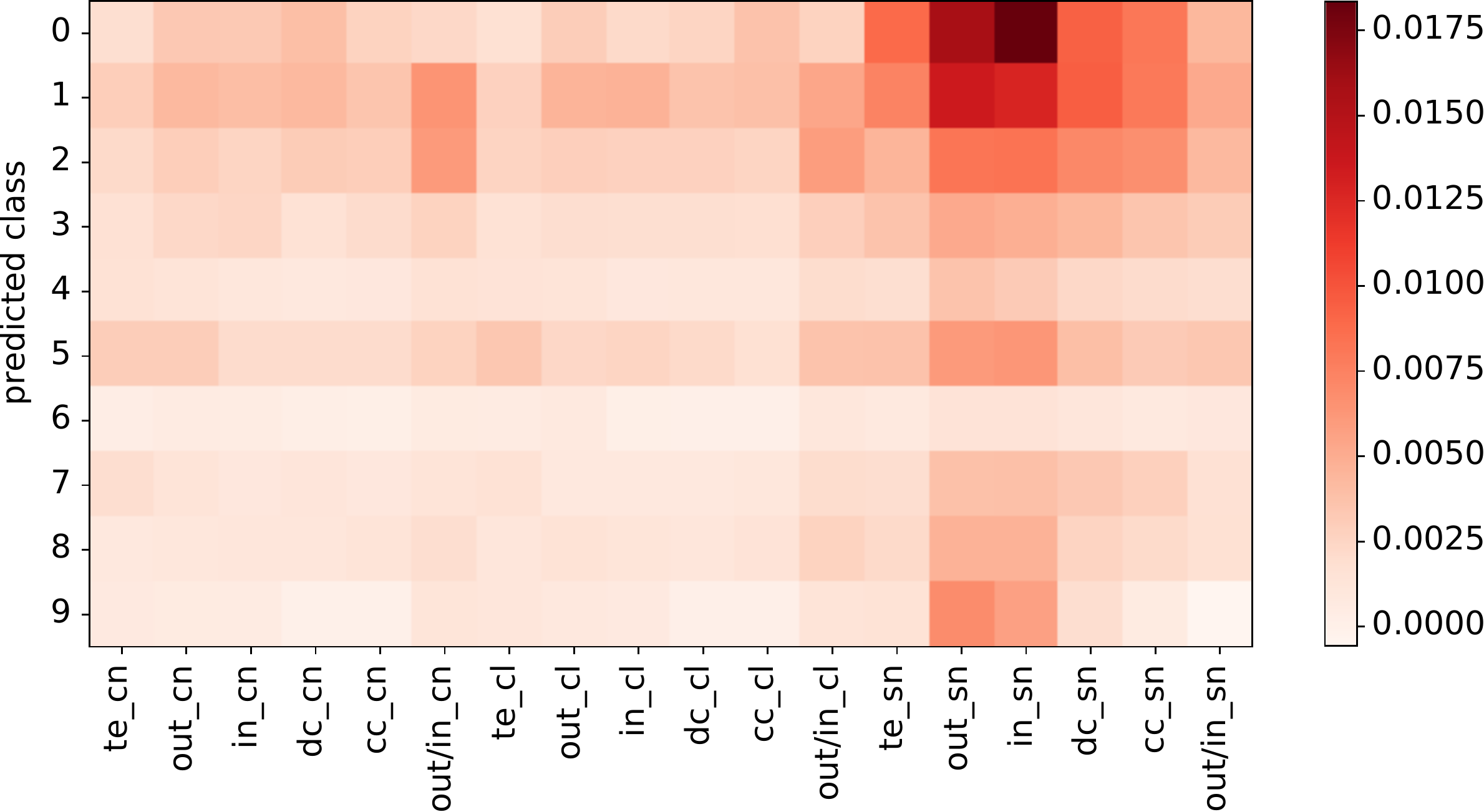}
    \caption{Explanations by sub-population, w.r.t. predicted class: for urban regions (left) and rural regions (right), based on LIME (top row) or SHAP (lower row).}
    \label{fig:explanations_disaggregated_}
\end{figure}

We also computed disaggregated explanations, depicted in 
figure \ref{fig:explanations_disaggregated_}.
They confirm the general relevance of features based on text message data for the prediction, as demonstrated already in the general case (previous section).
Unsurprisingly, they show a different distribution of classes: while urban areas belong to class 6 or higher, rural areas span all classes.
This is also reflected in the relevance of the features based on text message data, we observe the highest importance values, or contributions, as provided by LIME, or SHAP, respectively, for different classes, and depending on whether we look at urban or rural sub-populations.
We further observe that the feature importance values, or feature contributions, are consistently higher in magnitude for the urban compared to rural sub-population. This supports our finding that the model is biased w.r.t. the population density.

\begin{figure}[ht]
    \centering
    \includegraphics[width=0.475\linewidth]{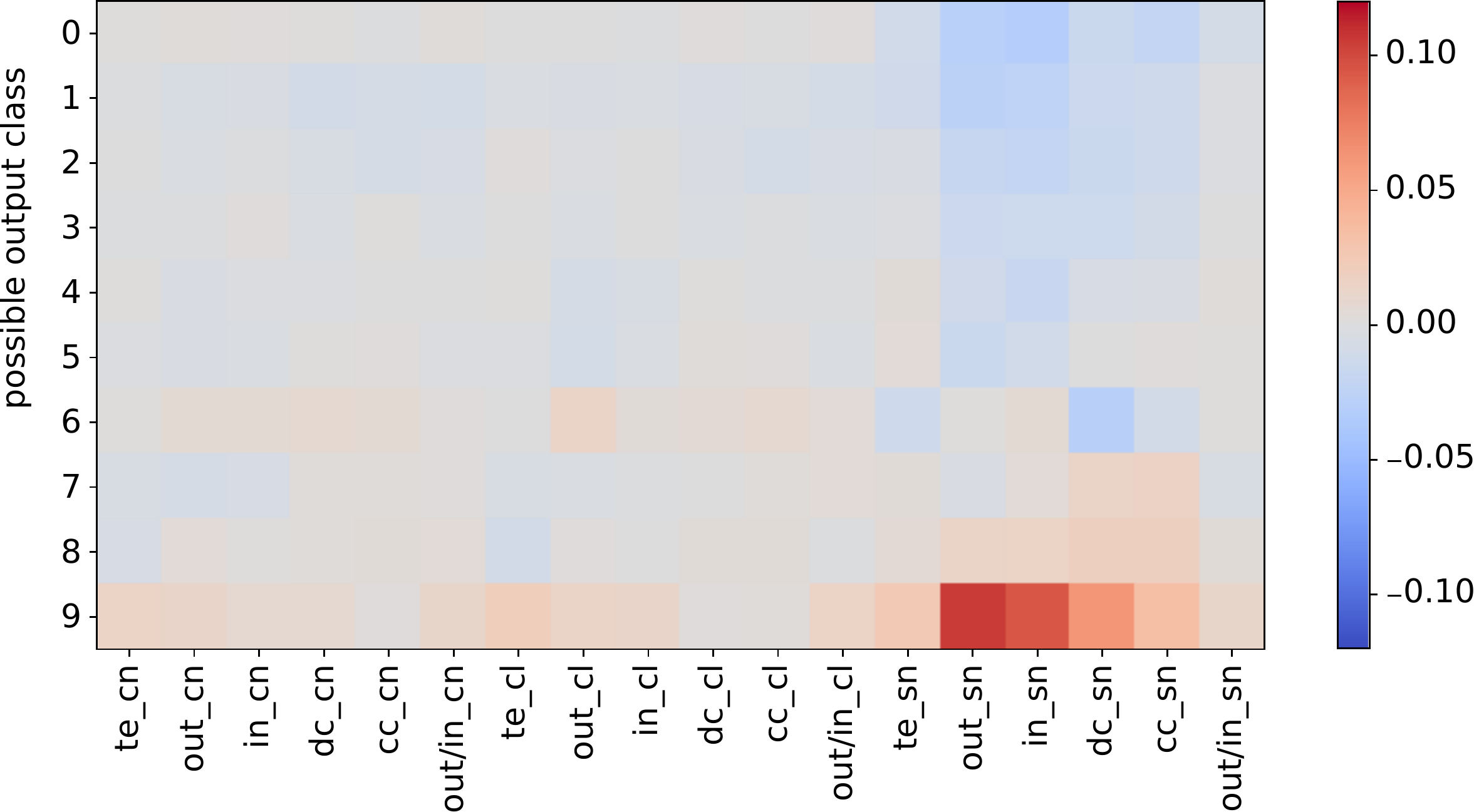}
    \includegraphics[width=0.475\linewidth]{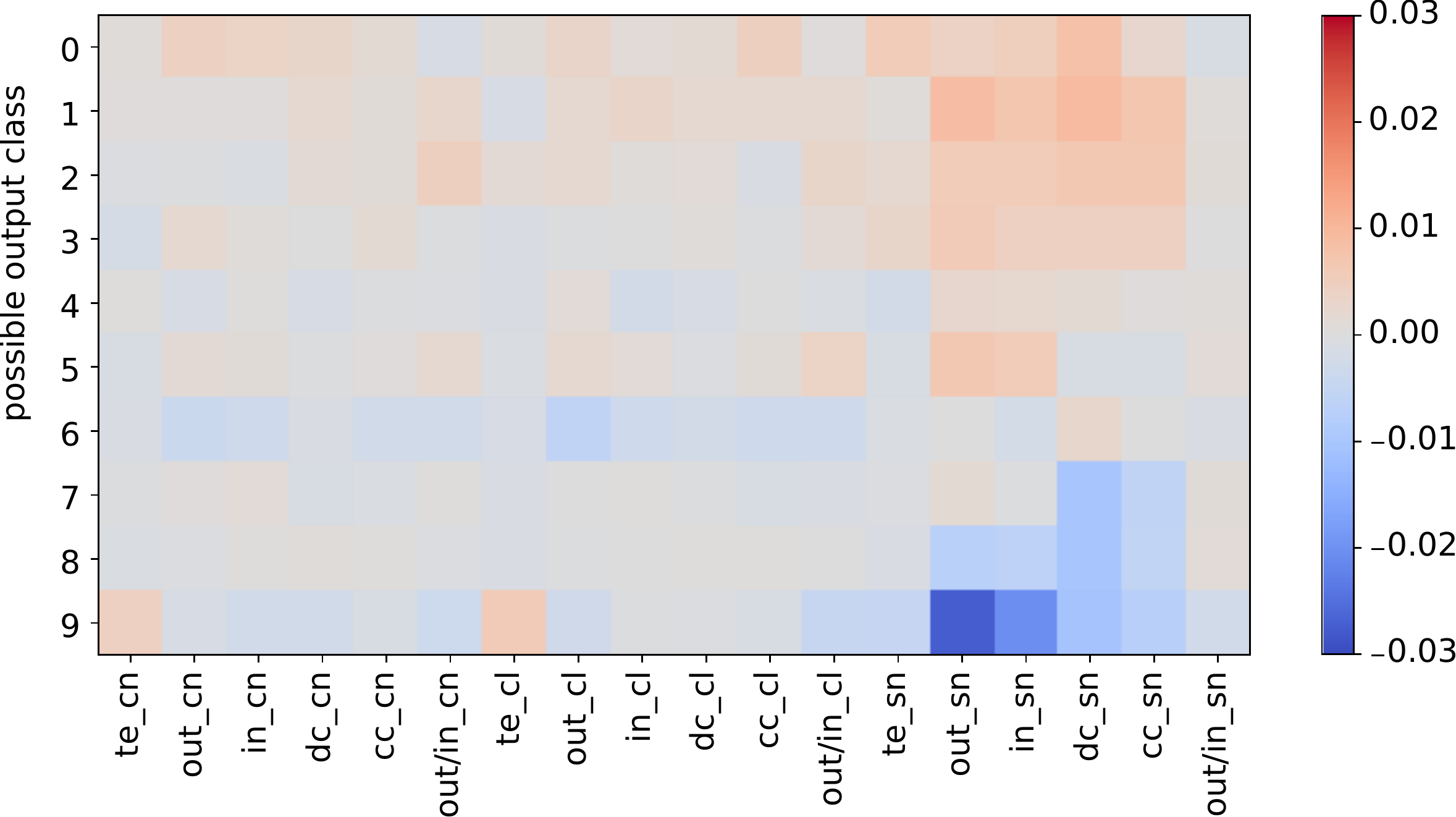}
    \caption{
    Explanations by sub-population, w.r.t. possible output classes as provided by SHAP: for urban regions (left), and rural regions (right).}
    \label{fig:explanations_disaggregated}
\end{figure}

In figure \ref{fig:explanations_disaggregated}, we show the explanations as provided by SHAP, w.r.t to \textit{all possible} output classes, and disaggregated by sub-population.
Similar to what we observed already in the previous section, we find in this case positive and negative values. 
As urban areas are tied to higher electrification classes, and rural areas to lower electrification classes, positive and negative values are correlated with opposing classes. Again, magnitudes are generally higher for urban compared to rural areas.

\section{Limitations}
\label{ref:limitations}

Although the data come from a network provider who is the market leader ($65\%$ in 2013), they are not fully representative of the population. 
A good starting point to investigate this further is the work by Pestre et al.~\cite{Pestre2019}.
Salat et al.~\cite{salat2020impact} point towards other biases, for example, due to a shift from the mobile phone network to relying increasingly on internet platforms such as Facebook.
Accounting for these biases, including further work on the bias w.r.t. population density is a next step. Also, we would like to apply other XAI methods such as LORE~\cite{DBLP:journals/corr/abs-1805-10820} to the ML model to understand whether we can extract some additional information.
{Another important extension is to run the explanations across all trained models and compare them with each other to understand better why there is a difference in performance.}

Data used in this work are proprietary and private. It is connected to individuals (their mobile phones) and to be processed only under their consent. It was provided in an anonymized form and further aggregated to safeguard individuals. Being proprietary, the data cannot be shared beyond the project. An alternative could be relying on other open-source data, such as satellite imagery, which have already been proven useful for similar projects. While being fundamentally different from mobile phone data, extending the XAI use-case to these data is a valuable path to follow.

The data we used originate from Senegal, a country of the Global South. Being situated in Western Europe, we should critically reflect on how this might perpetuate power relationships~\cite{DBLP:conf/fat/AbebeABKORS21}. While our focus is on providing a use-case of XAI, and to primarily verify a specific ML model that can estimate electrification rates in Senegal, we acknowledge the importance of local knowledge and domain expertise when evaluating data, and specifically when drawing policy implications~\cite{DBLP:conf/fat/AbebeABKORS21,DBLP:phd/us/Letouze16}.

\section{Conclusion}
\label{sec:conclusion}

In this work, after showing that electrification rates can be estimated from mobile phone data, we applied two local, model-agnostic explanation tools to verify our model. Both explanations perform well and agree with each other {on stressing the relevance of text message data for the predicted outcome. They confirm the general validity of the model.}
We also showed that our model is biased w.r.t. to population densities.
Thus, areas located in rural areas receive an unfair prediction, i.e. are more likely to be linked to a wrong electrification rate.

While the prediction of socio-economic indicators from remotely accessible data is not novel in itself, it is of high relevance, e.g., to support policy planning.
Verifying such a model using XAI techniques is certainly important, and novel. 
This is complementary to the fact that there are generally few use-cases applying an XAI method on a real-world problem, most of them centered around the Global North ~\cite{DBLP:conf/dev/OkoloDV22}. 

Our analysis showed that XAI methods can be useful to verify an ML model in practice. However, we would like to caution against using these tools blindly, and summarize the challenges that emerged during our work as follows:

\paragraph{Pipeline Design} If the aim is to use an XAI method, data processing and choosing the model tasks are limited by the fact that most XAI methods focus on tabular, image or text data, and on classification problems. Adapting a problem to this could lead to a loss of information and lower prediction accuracy.
Thus, efforts should be made to provide explanations for other task and data types (such as LASTS~\cite{DBLP:conf/cogmi/GuidottiMSPG20} for time series data, and beyond).

\paragraph{Domain Knowledge} While an isolated XAI method can be very useful for debugging purposes, domain knowledge is necessary to draw real-world connections, and eventually verify the model via the explanation.
In our work, an example of such knowledge is the distribution of text message activity over Senegal.
While domain knowledge needs to be available in the first place, it is usually external to the explanation. A direct integration into XAI methods via symbolic approaches could be therefore highly useful~\cite{DBLP:journals/corr/abs-2105-10172,DBLP:journals/ia/CalegariCO20}. 
\vspace{0.25cm} \\
The work presented in this paper relies on a static network. We initially kept a second version of the dataset in the form of time series. Details on these experiments, including the trained ML model and explanations based on LASTS~\cite{DBLP:conf/cogmi/GuidottiMSPG20} can be found in the appendix \ref{sec:lasts}. The time series data posed several challenges, among others, a high need of computing resources, few XAI approaches to use and compare, and a considerably higher amount of domain knowledge {required} for the interpretation of the data. For these reasons, we decided to continue working on the network data and leave the time series data for future work.

\section*{Acknowledgements}
Thanks to Salvatore Ruggieri and Franco Turini.
This work has received funding from the European Union’s Horizon 2020 research and innovation programme under Marie Skłodowska-Curie Actions (grant agreement number 860630) for the project ``NoBIAS - Artificial Intelligence without Bias'' (\href{https://nobias-project.eu/}{https://nobias-project.eu/}).
This work reflects only the authors’ views and the European Research Executive Agency (REA) is not responsible for any use that may be made of the information it contains.

\section*{Authors' Contributions} 
\textit{Laura State:} conceptualization, data processing and experiments, paper draft, writing and editing
\textit{Hadrien Salat:} data preparation, paper co-writing and editing
\textit{Stefania Rubrichi:} data curation, paper reviewing, co-supervision 
\textit{Zbigniew Smoreda:} data curation, paper reviewing

\newpage

\bibliographystyle{splncs04}
\bibliography{library.bib}

\newpage

\appendix
\section{Appendix}

\subsection{Data Distribution}
\label{sec:data_distribution}

\begin{figure}[ht]
    \centering
    \includegraphics[width=0.45\textwidth]{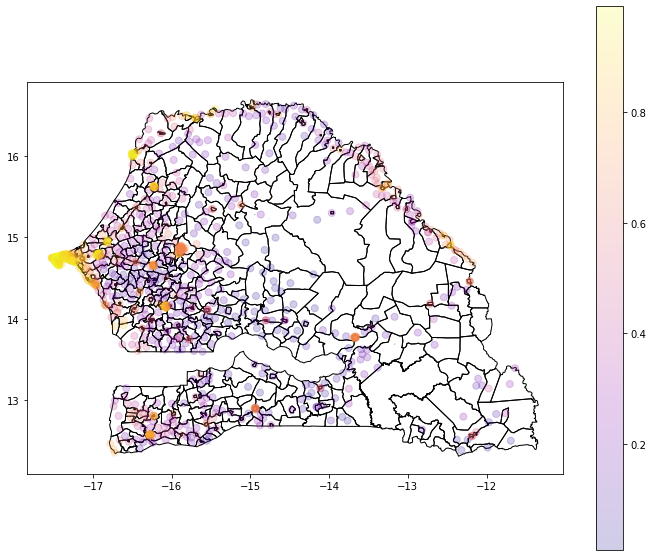}
    \includegraphics[width=0.45\textwidth]{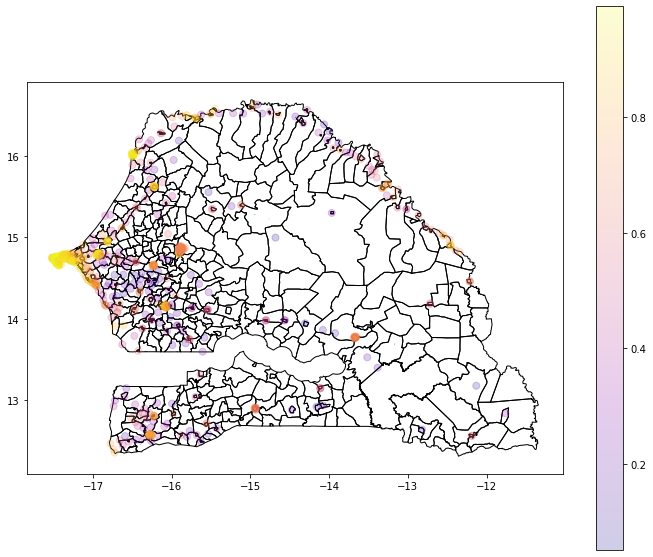}
    \caption{Spatial distribution of available data. Left panel: call data, right panel: text message data. Colored points reference to cell tower locations. Call data based on CN.
    Plots based on time series data, outgoing.}
    \label{fig:mpd_spatial_distribution}
\end{figure}

Figure \ref{fig:mpd_spatial_distribution} shows the distribution of available call data (left panel) and text message data (right panel).
While both data are more dense in the Western part of Senegal, text message data are specifically sparse in the Eastern part of the country.

\subsection{Additional Results}
\label{sec:additional_results_accuracies}

Additional results for the prediction of the electrification rate are shown in table \ref{tab:results2}.
This second set of models is also trained with default parameters.

\begin{figure}
    \centering
    \includegraphics[width=0.45\textwidth]{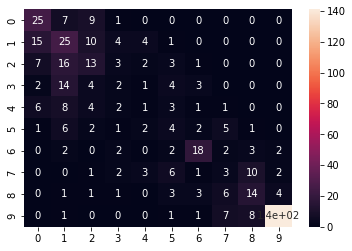}
    \caption{Confusion matrix for the random forest model (accuracy of $0.516$, MAE of $0.972$, best performing model.}
    \label{fig:confusion_matrix_rf}
\end{figure}

Further, we present a confusion matrix of the best performing model (RF, figure \ref{fig:confusion_matrix_rf}).

\bgroup
\begin{table}
    \centering
    \caption{Classification results. The higher the accuracy ($acc$) the better, the lower the $MAE$ the better. 
    }
    \begin{tabular}{|c|c|c|c|c|c|}
    \hline
    model & $acc$ & $MAE$ & $MAE$ / $MAE_{max}$ & $acc_{urban}$ & $acc_{rural}$ \\
    \hline
    \hline
         LOG & 0.463 & 1.331 & 0.148 & 0.778 & 0.264 \\
         ADA & 0.195 & 1.463 & 0.162 & 0.119 & 0.243 \\
         SVC & 0.465 & 1.296 & 0.144 & 0.784 & 0.264 \\
         MLP & 0.455 & 1.138 & 0.126 & 0.741 & 0.274 \\
         BAY & 0.294 & 1.604 & 0.178 & 0.384 & 0.236 \\
    \hline
    \end{tabular}
    \label{tab:results2}
\end{table}
\egroup

\subsection{Time Series Data}
\label{sec:lasts}

In this section, we briefly describe the work on time series data.

\subsubsection{Data Processing}

A time series $S = s_1 .. s_T$ consists of $T = 24 \times 12$ ordered data points, each being the monthly average of the aggregated number of events
per hour, such that $s_t, t \in 1 .. 24$ represent the monthly average of the aggregated number of events per hour in January (``daily activity curve '' for January), etc.
Events are separated by direction (incoming or outgoing). Thus, per cell tower, we create \textit{six} time series.
We refer to the TS dataset based on number of calls as CN, based on length of calls as CL and based on number of text messages as SN, and use out/in for outgoing/incoming activity, respectively.
We standardize each of the time series separately by applying the min-max scaler as provided by sklearn.
Data labeling and subsampling applies as above. Data partitioning follows~\cite{DBLP:conf/cogmi/GuidottiMSPG20}.

We find that text message data is heavily imbalanced, and that the dataset is smaller than the other datasets. Thus, we exclude this data from the time series analysis.

The variational autoencoder that is used in the explanation as displayed below, is trained for $k = 50$ dimensions and over $e = 500$ epochs.
We used the ``out CL'' data and model for explanations as it provides the smallest $MAE$.

\subsubsection{Classification}

To classify based on time series data, we use ROCKET (RandOm Convolutional KErnel Transform)~\cite{DBLP:journals/datamine/DempsterPW20}, a method based on random convolutional kernels for feature extraction and linear classification.
Results are displayed in table \ref{tab:rocket}. 

\begin{table}[]
    \centering
    \caption{Classification results. The higher the accuracy the better, the lower the $MAE$ the better.}
    \begin{tabular}{|c|c|c|c|}
    \hline
    data & $acc$ & $MAE$ & $MAE$ / $MAE_{max}$ \\
    \hline
    \hline
         out CN &  0.605 & 0.758 & 0.084  \\ 
         out CL &  0.600 & 0.637 & 0.071  \\ 
         in CN &  0.601 & 0.761 & 0.085  \\
         in CL &  0.532 & 0.814 & 0.090  \\
    \hline
    \end{tabular}
    \label{tab:rocket}
\end{table}

\subsubsection{Explanations}

\begin{figure}[h]
    \centering
    \includegraphics[scale = 0.25]{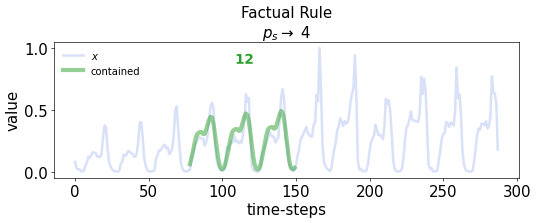}
    \includegraphics[scale = 0.25]{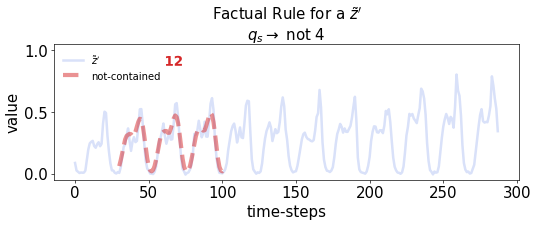}
    \caption{Local explanation by LASTS, shapelet-based, out CL data. Explained time series belongs to class 4, correctly classified by ML model. Left: factual rule, plotted against original time series, right: rule of opposite class, plotted against synthetically generated time series.
    Time steps in hours.
    }
    \label{fig:lasts1}
\end{figure}

Figure~\ref{fig:lasts1} shows a sample explanation by LASTS~\cite{DBLP:conf/cogmi/GuidottiMSPG20}. Explanations are provided in visual form and as rule, the latter can be read off the plot. 
The time series belongs to class 4, i.e. has an electrification rate between $0.4$ and $0.5$, and is correctly classified by the model. In the left panel, the factual rule is plotted against the original time series. The number above the shapelet indicates its index.
The rule reads as follows: ``If shapelet no. 12 is contained in the time series, then it is classified as class 4.'' This is mirrored by the rule for instances belonging to the opposite class (here: class 0..3, 5..9): `If shapelet no. 12 is not contained in the time series, then it is not classified as class 4.'', displayed in figure~\ref{fig:lasts1}, right, plotted against a synthetically generated time series from a class different to class 4.

\end{document}